\input{aipcheck}
\newcommand\araa{{ARA\&A}}%
\newcommand\apj{{ApJ}}%
\newcommand\apjl{{ApJ}}%
\newcommand\aap{{A\&A}}%
\newcommand\ssr{{Space~Sci.~Rev.}}%
\newcommand\jgr{{J.~Geophys.~Res.}}%
\def\kms{\hbox{km s$^{-1}$}}
\def\kapp{\hbox{$\kappa$}}
\def\bdotr{\hbox{B$\cdot$R}}
\def\cc{\hbox{cm$^{\rm -3}$}}
\def\glon{$\ell$}
\def\glat{$b$}
\def\elon{$\lambda$}
\def\elat{$\beta$}
\def\microG{\hbox{$ \mu{\rm G}$}}
\def\deeg{\hbox{$ ^\circ$}}

\documentclass[
    ,final            
    ,numberedheadings 
  ]
  {aipproc}

\layoutstyle{6x9}

\begin{document}

\title{How Local is the Local Interstellar Magnetic Field? }


\author{Priscilla C. Frisch}{
	address={University Chicago, Chicago, Illinois}
}

\begin{abstract}
Similar directions are obtained for the local interstellar magnetic
field (ISMF) by comparing diverse data and models that sample five
orders of magnetic in spatial scales.  These data include the ribbon
of energetic neutral atoms discovered by the Interstellar Boundary
Explorer, heliosphere models, the linear polarization of light from
nearby stars, the Loop I ISMF, and pulsars that are within 100--300
pc.  Together these data suggest that the local ISMF direction is
correlated over scales of $\sim 100$ pc, such as would be expected for
the interarm region of the galaxy.
\end{abstract}

\maketitle


\section{Introduction}

The properties of magnetic turbulence in interarm versus arm regions
are distinctly different, as shown by Faraday rotation measure (RM)
structure functions determined separately for the two regimes. The
interstellar magnetic field (ISMF) is coherent over 100 pc scales in
interarm regions, and random over spatial scales of less than 10 pc in
arm regions \cite{Haverkorn:2006}.  Which of these scenarios dominates
for the ISMF in the solar vicinity?  The heliosphere, and its ribbon of
energetic neutral atoms (ENAs) discovered by the Interstellar
Boundary Explorer (IBEX) spacecraft \cite{McComas:2009sci}, probe the
direction and strength of the very local ISMF in a single spatial
location.  Over intermediate scales, the position angles of starlight
polarized by magnetically aligned interstellar dust grains trace the
ISMF direction.  Pulsar RM data give the field direction for spatial
scales of 100--300 AU.  As discussed in Frisch et
al. \cite{Frisch:2011ismf2}, together these data and cosmic ray
anisotropies suggest the ISMF near the Sun is similar to expectations
for interarm regions.

\section{IBEX Ribbon, Heliosphere, and Local ISMF}\label{sec:ibex}

\begin{figure}[t!]
  \includegraphics[height=.3\textheight]{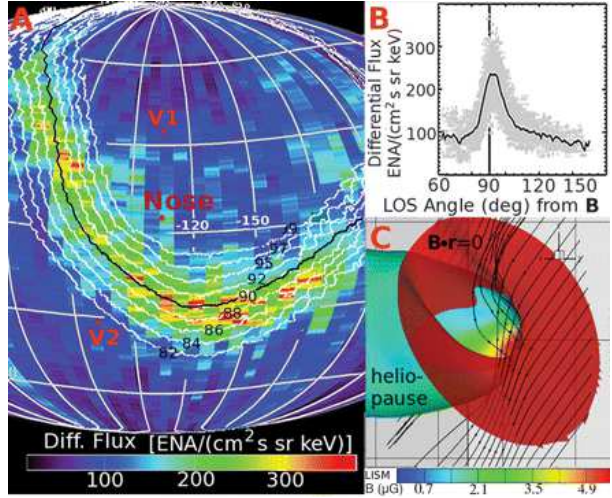}
  \caption{(A) Contours of constant angular offsets from the pole of
the IBEX ribbon at \elon=221\deeg, \elat=39\deeg, shown plotted
against the IBEX 1 keV ENA fluxes.  The black line shows the locus of
points perpendicular to the ISMF direction, e.g. \bdotr=0, based on
the ISMF displayed in (C) from the heliosphere model of
\cite{Pogorelovetal:2009asymmetries}.  B is the magnetic field vector
and R is the radial sightline. Plot (B) shows the 
correlation between global 1 keV ENA fluxes and \bdotr.  Plot (A) is in
ecliptic coordinates, and centered on the heliosphere nose.  The locations
of the two Voyager spacecraft are identified.  The figure is from
\cite{Schwadron:2009sci}.}
\label{fig:1}
\end{figure}

IBEX measures ENAs formed by charge-exchange (CEX) between
heliospheric ions and neutral gas from the surrounding ISM, which
flows through and around the heliosphere at a velocity of $\sim 26$
\kms\ from the direction \glon,\glat$\sim 4^\circ, 15^\circ$.  Among the sources of the ions creating the 0.01--6 keV ENAs
detected by IBEX are the post-shock solar wind, pickup ions, and
non-thermal energetic ions in the inner heliosheath, secondary ions
created by CEX between interstellar protons and ENAs escaping beyond
the heliopause, and possibly pickup ions inside of the solar wind
termination shock.  The first ENA full skymap, collected early 2009,
revealed an unexpected "ribbon" of ENAs that traces heliosphere
asymmetries created by the angle between the ISMF and velocity of
inflowing neutral interstellar gas
\cite[Fig. \ref{fig:1},][]{McComas:2009sci,Funsten:2009sci,Schwadron:2009sci}.
The ribbon is visible in the energy range $\sim 0.2 - 6 $ keV and
forms a nearly complete circle in the sky, that is offset by $\sim
75^\circ$ from the ribbon arc center
\cite{Schwadronetal:2011sep}.  Comparisons between heliosphere
models (see below) and the ENA ribbon revealed that the ribbon is seen
towards sightlines that are perpendicular to the ISMF draping over the
heliosphere \cite{Schwadron:2009sci}.  Seven possible formation
mechanisms for the ribbon have been discussed
\cite{McComasetal:2010var,Schwadronetal:2011sep,McComas:2009sci,GrzedzielskiBzowski:2010ribbon}.
In order to link the ribbon directly to the ISMF draping over the
heliosphere, the mechanism based on CEX between interstellar H$^\circ$
and energetic ions beyond the heliopause is assumed to be correct
\citep{Heerikhuisen:2010ribbon,Chalov:2010ribbon}.  The ribbon
location is stable over the first several skymaps
\cite{McComasetal:2010var}.  The ISMF direction at the heliosphere is
then given by the center of the ribbon arc, located at \glon=33\deeg,
\glat=55\deeg\ (see Table 1 for ecliptic coordinates).

The best match between the predicted and observed ribbon locations occurs for
ENAs produced roughly 100 AU beyond the heliopause
\cite{HeerikhuisenPogorelov:2011}.
The surrounding cloud is $\sim 22$\% ionized \cite[$n(\mathrm{p}) \sim 0.07$ \cc,][]{SlavinFrisch:2008}.  
The ribbon formation mechanism requires that ENAs escaping the heliosphere
CEX with interstellar protons piled up against the heliopause,
creating a secondary pickup ion population in the outer heliosheath.
The secondary pickup ion must CEX quickly with an interstellar H$^\circ$
atom (before magnetic turbulence disrupts the ring-beam distribution),
to create an ENA detected preferentially in directions perpendicular
to the ISMF.  The relative roles of large- and small-scale magnetic
turbulence in the outer heliosheath are not presently understood, so
this mechanism is still under study
\cite{Florinski:2010ribbon,GamayunovZhang:2010ribbon}.  A recent 
model 
\cite{HeerikhuisenPogorelov:2011} reproduces the ribbon for
an ISMF strength of $\le 3$ \microG, with the pole directed from
\glon=33\deeg, \glat=54\deeg\ (or $\lambda,\beta =
222^\circ,39^\circ$).  A local ISMF strength of $\sim 3$ \microG\ is
found from photoionizaton models of the local interstellar cloud (LIC)
around the heliosphere, and the assumption of pressure equilibrium
between the ISMF and thermal gas
\cite{SlavinFrisch:2008}.

MHD heliosphere models are constructed to reproduce the 
heliospere asymmetries induced by the $\sim 45^\circ$ angle between
the ISMF direction and gas flow direction
\citep[][]{Pogorelovetal:2009asymmetries,Prestedetal:2010,Opheretal:2007,Ratkiewiczetal:2008}.
The ISMF direction in these models is generally directed downwards
through the ecliptic plane in order to reproduce asymmetries such as
the $\sim 5^\circ$ offset between the central directions of the flows
of interstellar H and He into the heliosphere, and the difference
between the solar wind termination shock distances encountered
by Voyager 1 (94 AU) versus Voyager 2 (84 AU).

If the IBEX ribbon is formed beyond the heliopause through CEX with
secondary pickup ions, then the ribbon is an extraordinary diagnostic
of the local ISMF and neutral densities.  Models show that variations
in interstellar properties of $\sim 15$\% lead to pronounced
differences in the location and width of the ribbon because of the
interplay between magnetic, plasma and neutral pressures
\cite{Frisch:2010next,HeerikhuisenPogorelov:2011}.

Two aspects of the ribbon data are notable: (1) The imprint of the
solar wind on ENA fluxes and the ribbon is apparent in flux decreases
of $\sim 15$\% between the first and second skymaps, spaced by six
months \cite{McComasetal:2010var}.  The travel time for a 1 keV solar
wind proton to pass the Earth, the solar wind termination shock, and
then return to IBEX as an ENA, is over two years.  This time lag
suggests that the decrease in 0.7--1.1 keV fluxes in the second map is
from the unusually low solar wind pressure of the recent solar cycle
minimum.  (2) The energy distribution of global ENAs shows an overall
power law spectra, $E^{-\kappa}$, with \kapp\ ranging between 1.28 and
2.04 for different regions of the sky and spectral intervals
\cite{Funsten:2009sci}.  The power-law distribution of ribbon ENAs becomes softer above a
"knee" in the spectrum between 1--4 keV, which occurs at higher
energies for higher latitudes
\cite{Schwadronetal:2011sep}.

\section{Local ISMF direction from polarized starlight}

The galactic ISMF was first mapped in the 1940's using polarized
starlight, which showed patterns of linear polarization related to
dust extinction.  In the diffuse ISM, dust grain alignment occurs when
magnetic torques align asymmetric rotating charged dust grains,
producing a birefringent interstellar medium with lowest opacities parallel to the
ISMF direction. This gives linear polarization that is parallel to,
and traces, the ISMF direction \cite{Roberge:2004}. For sightlines
perpendicular to the ISMF direction in a uniform medium, polarization
will increase with distance.  Polarizations do not provide the
strength or polarity of the ISMF, and polarization is reduced by
foreshortening towards the poles of the field.
\begin{figure}[t!]
  \includegraphics[height=.3\textheight]{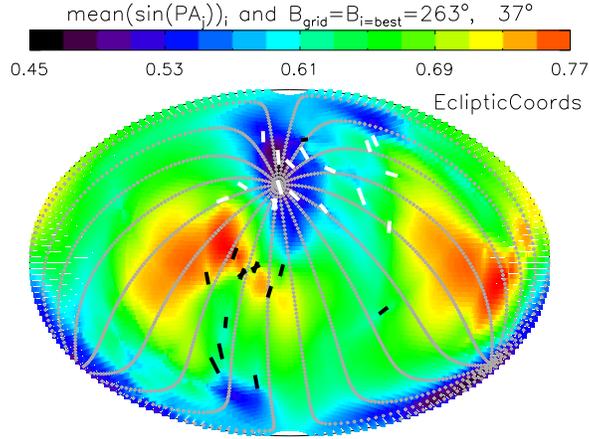}
\caption{Best best-fitting ISMF direction from polarization data is
determined by minimizing the function $< \sin(\mathrm{PA_i}) > $.
$\mathrm{PA_i}$ is the position angle of linear polarizations in
the i$^{\mathrm{th}}$ possible ISMF direction. The mean is
evaluated for data towards stars within 40 pc and 90$^\circ$ of the
heliosphere nose (where the G-cloud is seen, see
\cite{Frisch:2010ismf}).  Gray dots show the best fitting ISMF, which
is directed towards \elon=263\deeg, \elat=37\deeg, with uncertainties
of $\sim \pm 35^\circ$ (corresponding to \glon,\glat=38\deeg,23\deeg).
A new study using an extended polarization
data set returns similar results for the local ISMF direction, but the
uncertainties are larger \cite{Frisch:2011ismf2}.  Polarization
vectors are plotted with either black or white bars.
The figure is from \cite{Frisch:2010ismf}.}
\label{fig:2}
\end{figure}

During the early 1970's, Tinbergen 
\cite{Tinbergen:1982} conducted a sensitive (for that time) study of
the interstellar polarizations of stars within $\sim 40$ pc of the
Sun, and found a patch of dust in the south and close to the Sun.
Based on interstellar absorption lines towards $\alpha$ Cen at 1.3 pc,
and polarization towards 36 Oph, the magnetically aligned dust must be
in the "G-cloud" within 1.3 pc \cite{Frisch:2011araa}.  Recent
polarization data confirm that local interstellar polarizations are
significantly weaker in the northern hemisphere than the southern
hemisphere.

We have analyzed available interstellar polarization data, including
six recent sets of sensitive data, to determine the best fitting ISMF
direction for a 180\deeg\ diameter region that is centered on the
heliosphere nose and coinciding roughly with the G-cloud
(\cite{Frisch:2010ismf,Frisch:2011ismf2}.  Although we utilize
polarization towards stars out to 40 pc, most of the polarization
should be formed in the region with the highest concentrations of gas
within $\sim 15$ pc \cite{Frisch:2011araa,Frisch:2011ismf2}.  These
data indicate a local ISMF direction with a pole towards
\glon,\glat$\sim 38^\circ, 23^\circ$, with uncertainties of $\sim \pm
35^\circ$ (Fig. \ref{fig:2}, Table 1,
\cite{Frisch:2010ismf,Frisch:2011ismf2}).  This direction is centered
approximately 35\deeg\ from the center of the IBEX ribbon arc.  The
small difference between the two directions may indicate turbulence in
the local ISMF, for example between the ISMF directions of the G-cloud
(within 1.3 pc) and the LIC in which the Sun is embedded.  In several
regions, the flow of interstellar gas past the Sun shows merging or colliding
clouds \cite{RLIV:2008vel}, such as the two clouds in front of Sirius
\cite{FrischMueller:2011ssr}, which could drive local magnetic
turbulence.

\begin{table}[b!]
\begin{tabular}{ccc cc}
\hline
  \tablehead{1}{c}{b}{Scales}
& \tablehead{1}{c}{b}{ISMF Source} 
& \tablehead{1}{c}{b}{Angle} 
& \tablehead{1}{c}{b}{\elon, \elat}
& \tablehead{1}{c}{b}{\glon,\glat}  \\
\hline
$\sim 150 $ AU & IBEX ribbon center \cite{Funsten:2009sci} & 0\deeg & 221\deeg,
39\deeg & 33\deeg, 55\deeg \\ 
$\sim 150 $ AU & Heliosphere model \cite{HeerikhuisenPogorelov:2011}\tablenote{The ISMF polarity is directed
downwards through the ecliptic plane} & 3\deeg & 224\deeg, 41\deeg &
36\deeg, 53\deeg \\ 
$<40$ pc & Optical polarization \cite{Frisch:2010ismf,Frisch:2011ismf2}  &
  32\deeg & 263\deeg , 37\deeg & 38\deeg, 23\deeg\ ($\pm35^\circ$) \\ 
 80 pc & S1 Magnetic Bubble \cite{Wolleben:2007}  & $46^{+26} _{-22} $ & 291\deeg,
	   64\deeg & 71\deeg, 18\deeg ($\pm 48^\circ$)\\ 
150 pc & Pulsar RM data \cite{Salvati:2010}\tablenote{The ISMF polarity is directed upwards through the ecliptic plane} & 23\deeg & 232\deeg, 18\deeg &
	   5\deeg, 42\deeg \\ 
\hline
\end{tabular}
\caption{Interstellar Magnetic Field Direction, 150 AU -- 150 pc }
\label{tab:one}
\end{table}

\section{The Loop I superbubble and local ISMF}

Loop I is an evolved but reheated superbubble that dominates the
northern hemisphere because of it's large angular radius. It appears as
a distinct imprint on the ISMF within 100 pc that is detected in polarized
starlight, polarized radio continuum, Faraday rotation, and Zeeman
splitting in the HI filaments defining the shell circumference 
\cite{Frisch:1995rev}.  All existing spherically symmetric models of the radio
continuum Loop I place the Sun in the superbubble rim.  A recent study
of the intensity of 1.4 GHz emission in this region derived two
subshells of Loop I containing enhanced synchrotron emissivity from
swept up magnetic fields, and placing the Sun in the rim of the S1
subshell \cite{Wolleben:2007,Frisch:2010s1}.  The direction of the
bulk flow of the "cluster of local interstellar clouds" (CLIC) past
the Sun has an upwind direction of
\glon,\glat=335\deeg,--7\deeg\  (\cite{FrischMueller:2011ssr}, after correcting for the solar
velocity through the local standard of rest, LSR),
near the center of the S1 shell,

If the CLIC is part of the nearside of an evolved superbubble shell
associated with Loop I \cite{Frisch:1995rev,Frisch:2011araa}, then
both the polarity and direction of the local ISMF could be ordered
over the $\sim 78$ pc radius of the S1 subshell.  The angle between
the local ISMF direction and the LSR CLIC velocity is
68\deeg--78\deeg, depending on whether the comparison is with the
polarization axis or ribbon arc center.  This large angle suggests
that the very local ISMF is roughly parallel to the local surface of
the expanding Loop I shell.

\section{ISMF from pulsars in the Local Bubble }

The ratio of pulsar Faraday rotation measures (RM) and dispersion
measures provides an electron-density weighted measurement of the ISMF
component parallel to the sightline, and the field polarity.  Salvati
(2010) performed a best-fit to the rotation and dispersion measures of
four pulsars within 160--300 pc, and in the very low density third
galactic quadrant corresponding to the interior of the Local Bubble.
He found a best-fit direction of
\glon$\sim5^\circ$, \glat$\sim 42^\circ$, and a field strength of
$|B|$=3.3 \microG\ (Table 1).  This direction is within 23\deeg\ of
the center of the IBEX ribbon arc, and 33\deeg\ of the best-fitting
ISMF direction.  The polarity of this ISMF is directed towards the
pole of \elon,\elat=232\deeg,18\deeg, indicating that it is directed
upwards through the plane of the ecliptic.\footnote{Note that the
rotation measure is positive when the ISMF is directed towards the
observer, by definition.}  Wolleben \cite{Wolleben:2007} finds a
similar RM polarity in this direction, although variations over small
angular scales are seen.

\section{Galactic cosmic rays and the local ISMF}

An unusual indicator of the ISMF affecting the heliosphere is provided
by GeV--TeV galactic cosmic rays (GCR).  Cosmic rays at sub-TeV
energies have gyroradii less than several hundred AU in a 1 \microG\
magnetic field, which is comparable to heliosphere dimensions.  GCRs
with energies $\le 10^3$ GeV show pronounced spatial asymmetries
\cite{Nagashimaetal:1998,Halletal:1999gcr,LazarianDesiati:2010},
including a sidereal distribution with a southern GCR deficit, and a
broad excess emission region attributed to the heliotail.  The broad
tail-in excess at 500--1000 GeV is centered at ecliptic coordinates of
\elon,\elat\ of 71\deeg,--42\deeg\ to 90\deeg,--53\deeg, and is skewed
with respect to ecliptic latitude and covers the direction opposite to
the ISMF defined by the ribbon arc (\elon,\elat=41\deeg,--39\deeg,
Table 1).  This tail-in excess also coincides with the axis of the
ISMF determined from the polarization data,
\elon,\elat=83\deeg,--37\deeg.  The minimum in tail ENA fluxes
is 44\deeg\ west of the downwind gas-flow direction
\cite{Schwadronetal:2011sep}, and it overlaps the excess of sub-TeV
particles.  Evidently the GCR excess could form either in deep tail
regions or indicate GCRs streaming along the local ISMF or S1 shell.
Lazarian and Desiati (2010) have modeled the tail-in sub-TeV excess
\nocite{LazarianDesiati:2010} 
as due to the stochastic acceleration of particles in the
time-variable magnetic field of the heliotail.  

\section{Summary}

The summary of data in the previous sections shows that the direction
of the ISMF appears to be relatively constant over scales of five
orders of magnitude, with warping (large-scale turbulence) or
variations in the field direction typically less than
30\deeg--40\deeg\ (Table 1).  The agreement between ISMF direction
from pulsars in the Local Bubble, starlight polarizations, the IBEX
ribbon, and heliosphere models, is remarkable considering that the
spatial scales of these estimates differ by five orders of magnitude.
The puzzle is that opposite polarities are inferred for the ISMF
shaping the heliosphere and the pulsar RM data, downwards through
the ecliptic plane in the first case and upwards through the ecliptic
plane in the second case.  It could be pure coincidence that the Local
Bubble ISMF direction from the pulsar data and from heliosphere
measurements and models are similar.  However the galactic magnetic
field is known to be ordered over kiloparsec spatial scales in low
density interarm regions of the Galaxy such as around the Sun
\cite{JinLinHan:2009}.  Perhaps it is not so surprising to see
uniformity of the magnetic field direction over local scales of 100
pc.  If the ISMF traced by the IBEX ribbon, polarization data, Loop I
shell, and pulsar data are related, it would mean that Loop I is a
highly asymmetric evolved superbubble, with the Sun in the segment
most distant from the superbubble source and most closely linked to
the interarm field direction.


\begin{theacknowledgments}
The author thanks Paolo Desiati for pointing out the usefulness of
cosmic ray TeV asymmetries for understanding the relation between the
heliosphere and local ISMF.  This work was supported through grant
NNX09AH50G to the University of Chicago, and by the Interstellar
Boundary Explorer Mission as a part of NASA's Explorer Program.
\end{theacknowledgments}



\end{document}